\documentclass[conference]{IEEEtran}
\IEEEoverridecommandlockouts
\pdfoutput=1 
\usepackage{cite}
\usepackage{amsmath,amssymb,amsfonts}
\usepackage{algorithmic}
\usepackage{graphicx}
\usepackage{textcomp}
\usepackage{xcolor}

\usepackage[T1]{fontenc}% optional T1 font encoding

\usepackage{amsmath}
\usepackage{paralist}
\usepackage{tabularx}
\usepackage{graphicx}
\usepackage{sidecap}
\usepackage{epstopdf}
\usepackage{amsthm}
\usepackage{textcomp}
\usepackage{cite}
\usepackage{ragged2e}

\usepackage[utf8x]{inputenc}
\usepackage[T1]{fontenc}
\usepackage{textcomp}

\DeclareUnicodeCharacter{8289} {\textdegree}

\def\BibTeX{{\rm B\kern-.05em{\sc i\kern-.025em b}\kern-.08em
    T\kern-.1667em\lower.7ex\hbox{E}\kern-.125emX}}
\begin{document}

\title{Learning-based Resource Optimization in Ultra Reliable Low Latency HetNets\\
{}
\thanks{Identify applicable funding agency here. If none, delete this.}
}

\author{\IEEEauthorblockN{1\textsuperscript{st} Mohammad Yousefvand}
\IEEEauthorblockA{\textit{Winlab},
\textit{Rutgers University},
NJ, US \\
my342@winlab.rutgers.edu}
\and
\IEEEauthorblockN{2\textsuperscript{nd} Kenza Hamidouche}
\IEEEauthorblockA{\textit{Winlab},
\textit{Rutgers University},
NJ, US \\
kenzaham@winlab.rutgers.edu}
\and
\IEEEauthorblockN{3\textsuperscript{rd} Narayan~B.~Mandayam}
\IEEEauthorblockA{\textit{Winlab},
\textit{Rutgers University},
NJ, US \\
narayan@winlab.rutgers.edu}
\and

}

\maketitle

\begin{abstract}
In this paper, the problems of user offloading and resource optimization are jointly addressed to support ultra-reliable and low latency communications (URLLC) in HetNets. In particular, a multi-tier network with a single macro base station (MBS) and multiple overlaid small cell base stations (SBSs) is considered that includes users with different latency and reliability constraints. Modeling the latency and reliability constraints of users with probabilistic guarantees, the joint problem of user offloading and resource allocation (JUR) in a URLLC setting is formulated as an optimization problem to minimize the cost of serving users for the MBS. In the considered scheme, SBSs bid to serve URLLC users under their coverage at a given price, and the MBS decides whether to serve each user locally or to offload it to one of the overlaid SBSs. Since the JUR optimization is NP-hard, we propose a low complexity learning-based heuristic method (LHM) which includes a support vector machine-based user association model and a convex resource optimization (CRO) algorithm. To further reduce the delay, we propose an alternating direction method of multipliers (ADMM)-based solution to the CRO problem. Simulation results show that using LHM, the MBS significantly decreases the spectrum access delay for users (by $\sim$ 93\%) as compared to JUR, while also reducing its bandwidth and power costs in serving users (by $\sim$ 33\%) as compared to directly serving users without offloading.
\end{abstract}


\section{Introduction}
The emergence of delay-sensitive applications such as intelligent transportation systems, and patient monitoring applications makes it necessary to redesign classical resource allocation techniques in wireless heterogeneous networks (HetNets) and support ultra-reliable and low latency communications (URLLC) \cite{Bennis:2018:URLLC}. URLLC introduces new challenges to the design of next-generation cellular networks where the traffic consists mainly in short packet transmission and the related hard constraints in terms of latency and reliability. In fact, any delay in the transmissions of the order of microseconds could make the packets useless and hence must be dropped. Coupled with the ultra-density of future cellular networks that are expected to support billions of Internet of things (IoT) devices, time-sensitive applications will require a large amount of network resources such as power and bandwidth. Thus, the optimization of such scarce resources represents a crucial challenge for wireless service providers, as they need to support URLLC in one hand, and reduce their cost in using network resources on the other.

Several works in the literature have addressed the problem of resource allocation in cellular networks for both bandwidth-intensive applications and URLLC traffic \cite{Chen:2017:DSA,Sun:2017:EER,Sutton:2018:EUR,Kasgari: 2018: SOC}. Most of these works have considered a cellular network model with a single base station (BS) that serves two types of users namely eMBB and URLLC users, and proposed techniques to jointly satisfy the delay and reliability constraints of URLLC users, while optimizing the allocation of resources for the cellular BS. The authors in \cite{Chen:2017:DSA} proposed an optimal resource allocation strategy for uplink transmissions to maximize the delay-sensitive area spectral efficiency as a performance metric while guaranteeing the constraints on reliability. In \cite{Sun:2017:EER}, the authors proposed a method for maximizing energy efficiency for URLLC under strict QoS constraints on both end-to-end delay and overall packet loss. In \cite{Sutton:2018:EUR}, the authors investigated the potentials of using unlicensed spectrum for enabling ultra reliable and low latency communications. The work in \cite{Kasgari: 2018: SOC} presented a network slicing based resource allocation framework to provide reliable and low latency communications to users with such demands.

Although interesting, all these works consider a network composed of a single cell while currently deployed networks are heterogeneous with different types of base stations. Thus, none of these works have studied the opportunity of offloading users to potential small cells as a possible way for increasing reliability of the transmissions. Moreover, they do not account for the impact of the serving cost on the allocation of resources and offloading at the service providers. User offloading and resource allocation are two effective and highly correlated techniques for enabling URLLC in wireless HetNets, and due to their interplay, they must be jointly optimized while considering the monetary impact on the service providers.

The main contribution of this paper consists in jointly considering user offloading and resource optimization to enable URLLC in HetNets. In our model, hard latency and reliability constraints of URLLC users are modeled with probabilistic guarantees and relaxed based on Markov's inequality.  In particular,  we formulate the joint user association and resource optimization (JUR) problem as an optimization problem which is NP-hard and computationally intractable for large HetNets.
We reduce the complexity of the JUR problem by casting it into two sub-problems and then proposing an efficient learning-based heuristic method (LHM) to solve them. The first sub-problem is the user-cell association problem for which we reformulate it as a classification problem and solve it using a support vector machine (SVM)-based learning algorithm. We use the results of JUR optimization to train the SVM classifier, and since the results of JUR problem are optimized, the training error will be minimized this way. Once we trained the SVM, we can use it to determine user associations for future users. The second sub-problem is the resource allocation problem for which we propose a low complexity iterative algorithm. After solving user association problem using the trained SVM classifier, the JUR problem will be simplified by removing binary user association variables from it and it will be reduced to a convex optimization problem. To solve such convex optimization problem for the MBS, we propose a low complexity iterative solution based on the alternating direction method of multipliers, in which we use a penalized Lagrangian function with a barrier penalty function for reliability constraints of URLLC users to make sure such constraints are satisfied. Simulation results show that the proposed heuristic method significantly decreases the spectrum access delay for users (by $\sim$ 93\%) as compared to JUR problem, while maintaining the full service rate. By offloading 75\% of users to SBSs, it also reduces the MBS's bandwidth and power consumption costs in serving users by 33\% as compared to to the method with no offloading referred to as the Direct Serving Method (DSM).
The rest of this paper is organized as follows. Section \ref{sec2} introduces the network and system models, and the JUR optimization problem is formulated in section \ref{sec3}. Section \ref{sec4} presents the proposed learning-based heuristic method which includes a SVM-based user association model and a low complexity iterative algorithm for resource optimization. Simulation results are presented in section \ref{sec5}, and we conclude this paper in section \ref{sec6}.

\section{Network and System Models}
\label{sec2}

\subsection{Network Model}
We consider a HetNet model which includes one macro cell BS (MBS) and several overlaid small cell BSs (SBSs), and a mix of users with different URLLC applications who are randomly distributed under the coverage area of the MBS. We assume that the MBS is primarily responsible for serving all users, however it can offload some users to overlaid SBSs if such BSs offer to serve users located under their coverage with a price which is less than the cost of serving them directly by the MBS. Figure below shows our network model, in which MBS uses high power transmissions (denoted with gray links) to serve users who are located in the cell edge boundaries, while SBSs can serve such users who are located under their coverage area with low power transmissions (denoted with yellow links), and hence with less cost.
\begin{figure}[htb!]
	\includegraphics[width=\linewidth]{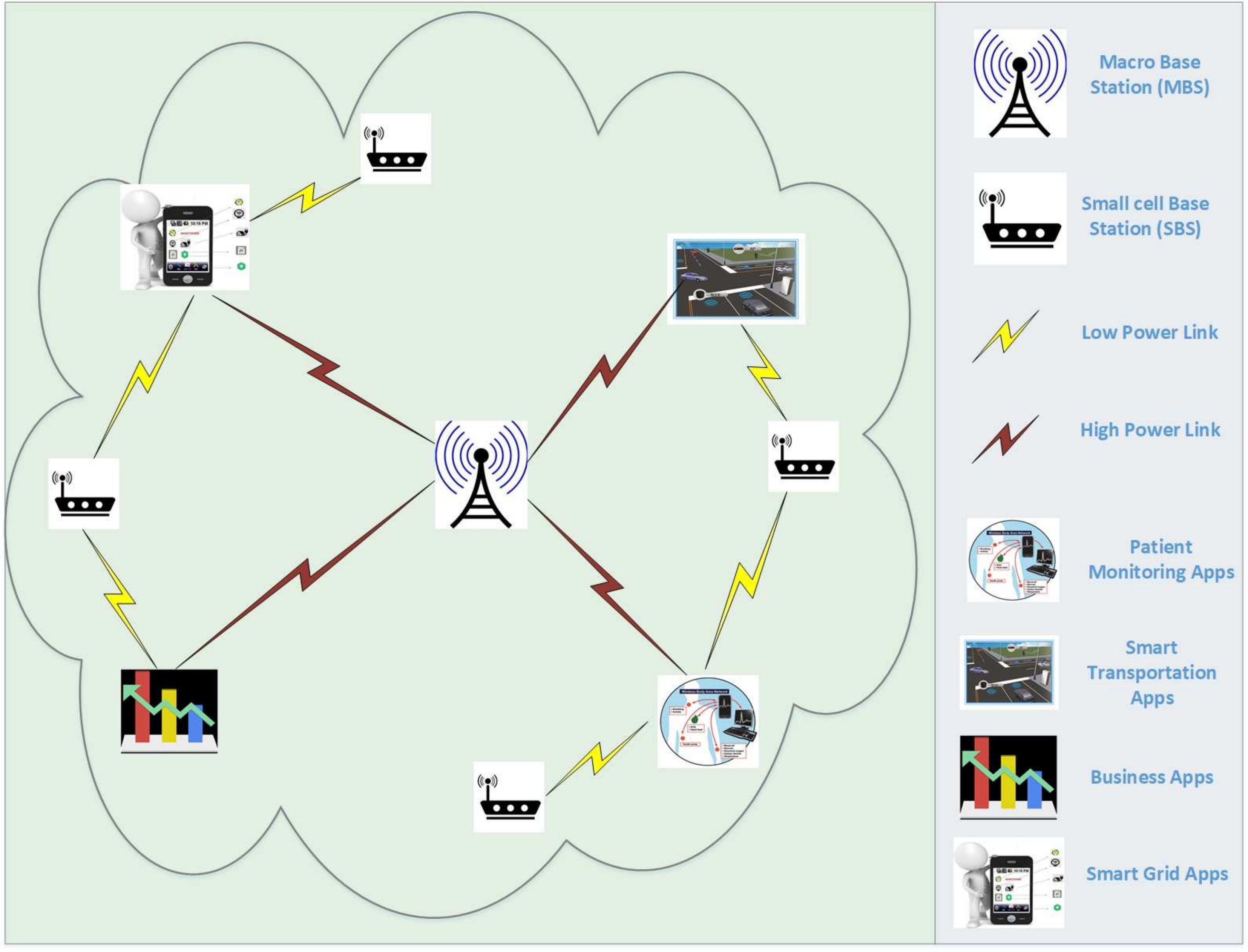}
	\centering
	\caption{HetNet Model with URLLC Applications.}
	\label{fig:boat1}
\end{figure}

\subsection{System Model}
In our system model, we assume that MBS periodically optimizes its decisions on user offloading and resource allocations, and at the beginning of such optimization intervals an auction will happen between MBS and SBSs, in which each SBS $k$ bids to serve each user $i$ under its coverage by calculating the cost of serving such user considering its delay, reliability and data rate constraints. We assume in each transmission request, each user includes its requirement in terms of latency, data rate and reliability which could be different from other user's requirements. The total price offered by the SBS $k$ to serve user $i$ is defined as $\Phi_{k,i}$, which is given by
\begin{equation}
\label{eq1}
\Phi_{k,i}=\Phi_{k,i}^s+ \Phi_{k,i}^r,
\end{equation}

\noindent as the summation of two terms; the first term $\Phi_{k,i}^s$ is the cost of bandwidth and power resources used by SBS $k$ to serve user $i$ and satisfy its constraints, and the second term $\Phi_{k,i}^r$ is the amount of reward asked by SBS $k$ to serve user $i$. This offered price has to be paid by the MBS to the SBS $k$, if MBS offloads user $i$ to the SBS $k$. The objective function of the MBS is to minimize its overall cost in serving and offloading all users in the HetNet, and is given by
\begin{equation}
\label{eq2}
\min _{(\mu_i, w_i, p_i)} \sum \limits_{i\in U} \mu_i(c_p p_i + \gamma c_w w_i) + (1-\mu_i)\Phi_{k,i}^*,
\end{equation}

\noindent in which $p_i$ and $w_i$ denote the amounts of power and bandwidth required by the MBS to serve user $i$, respectively. And $\mu_i$ is the binary user association variable for user $i$, with $\mu_i=1$ if users $i$ is associated to the MBS, and $\mu_i=0$ if user $i$ is offloaded to the best serving SBS who offers the minimum price to serve this user among all other SBSs. Also, $c_p$ and $c_w$ are the MBS unit costs for power and bandwidth, respectively, $\gamma$ is the regularization parameter which models the trade-off between power and bandwidth costs, $U$ denotes the set of all users, and $\Phi_{k,i}^*$ is the price offered by the best serving SBS to serve user $i$ and is given by
\begin{equation}
\label{eq3}
\Phi_{k,i}^*= \min _{k} \Phi_{k,i}.
\end{equation}

We assume the overall spectrum access delay for each user $i$, $d_i$, is given by
\begin{equation}
\label{eq4}
d_i= d_c + \mu_i d_o,
\end{equation}

\noindent which is the summation of MBS computation delay $d_c$, and offloading delay $d_o$ in case user $i$ is offloaded. So, $d_c$ accounts for the delay in making user associations decisions by MBS, and is a function of the computational complexity of the optimization problem used by MBS for user association and resource allocation, and its processing power, hence it is assumed to be fixed for all users. However, offloading delay is only considered for offloaded users and is assumed to be equal to $3*RTT$ as three $RTT$ is required for transmissions of bidding, bid selection, and acknowledgement messages between MBS and selected SBS. 

Denoting $r_{i}$ as the service rate of user $i$, it is given in
\begin{equation}
\label{eq5}
r_{i}= w_i ~log (1+ p_i {h_i}^2 /N_0).
\end{equation}

\noindent as a function of allocated power, $p_i$, and bandwidth, $w_i$, to this user, and its channel gain, $h_i$. 
We assume each user $i$ has a threshold for its acceptable delay, denoted as $d_{th,i}$, and the delay constraint for each user is defined by setting an upper bound, $\delta_d$, for the violation probability of its delay constraint as defined in 
\begin{equation}
\label{eq6}
Pr [d_{i} \geq d_{th,i}] \le \delta_d.
\end{equation}

Also, the data rate constraint for each user $i$, can be defined as the probability of satisfying its requested data rate, $r_{th,i}$,
\begin{equation}
\label{eq7}
Pr [r_{i} \geq r_{th,i}],
\end{equation}

and, the reliability constraint for each user $i$ is defined by setting an upper bound for its data rate constraint's violation probability,
\begin{equation}
\label{eq8}
Pr [r_{i} \leq r_{th,i}] \le \delta_r.
\end{equation}

\section{ Joint User Offloading and Resource Optimization (JUR)}
\label{sec3}
After defining the objective function for MBS in (\ref{eq2}), and delay and reliability constraints for users in (\ref{eq6}), and (\ref{eq8}), respectively, the joint user association and resource optimization (JUR) problem for the MBS can be formulated to minimize the cost of serving users for the MBS while satisfying their delay and reliability constraints. The JUR problem formulation is given by
\begin{subequations}	
	\begin{flalign}
	&\min _{(\mu_i, w_i, p_i)} \sum \limits_{i\in U} \mu_i(c_p p_i + \gamma c_w w_i) + (1-\mu_i)\Phi_{k,i}^*, &\label{eq9a},\\\nonumber\\
	&~~~~~ subject~to: \nonumber &\\
	&Pr [d_{i} \geq d_{th,i}] \le \delta_{d,i},~\forall i\in U, &\label{eq9b}\\
	& Pr [r_{i} \leq r_{th,i}] \le \delta_{r,i},~\forall i\in U, &\label{eq9c}\\
	& 0 \leq p_i \leq P_{max},~\forall i\in U,&\label{eq9d}\\
	& 0 \leq \sum \limits_{i\in U}w_i \leq W_{max},&\label{eq9e}\\
	& \mu_{i} - \mu_{i}^2 =0,~\forall i\in U,&\label{eq9f}\\
	\nonumber
	\end{flalign}
\end{subequations}

\noindent in which $p_{max}$ is the maximum power spectral density that can be used by MBS to serve any user, and $W_{max}$ is the total bandwidth available at the MBS. The constraints in (\ref{eq9d}) and (\ref{eq9e}) ensure that the allocated power and bandwidth to each user is within the acceptable range for them, respectively, and the constraint in (\ref{eq9f}) ensures that user association variable for each user $i$ is a binary integer variable. Solving the optimization problem defined in (\ref{eq9a})-(\ref{eq9f}) gives the optimal solution to the JUR problem, however this is a binary integer non-linear programming problem, which is NP-hard and computationally intractable for HetNets with large number of users. In fact, in \cite{{Yousefvand:2017:DES}} we showed that a simplified version of this problem is reducible to the Knapsack problem which is well known NP-hard problem, thus JUR optimization is also NP-hard and not scalable for large HetNets. Hence, we need to find low complexity alternative solutions to the JUR problem.

\section{Proposed Heuristic Method}
\label{sec4}
To increase the efficiency of resource allocation for the MBS, we replace the NP-hard JUR problem with a two phase low complexity heuristic solution, in  which we first solve the user association problem using a Support Vector Machine(SVM)-based user association (SUA) model, and then we optimize the MBS's power and bandwidth allocation using a gradient decent-based resource allocation (GRA) algorithm.

\subsection{SVM-based User Association (SUA)}
Since the user association variables in JUR optimization $(\mu_i, \forall i\in U)$ are binary variable and optimization problems with binary variables are often NP-hard, in this section we propose a learning based heuristic solution to user association problem, to remove such variables from JUR optimization problem. In fact, since MBS's decisions on user association fo all users are binary, to either serve them or offload them to SBSs, the user association problem in HetNets can be seen as a classification problem, to classify users between MBS and SBSs, which  can be efficiently solved for large HetNets using SVMs. Assuming we have the training data from running the JUR problem by MBS in previous time slots, we can train an SVM to lean the user association model from them, and use the trained SVM to predict the user association for the future time slots. To do so, we assume we have a set of labeled data points $(u_i, \mu_i)$ in which $\mu_i$ is the user association value for user $i$, and $u_i=({x_i, d_{th,i}, r_{th,i}, \delta_{d,i}, \delta_{r,i}, SNR_i})$ is the user i-th features vector which includes its distance to MBS, data rate threshold, reliability threshold, data rate violation bound, reliability violation bound, and the SNR of its signal at the MBS. The training set $D$ which includes $N$ data points is given by
\begin{equation}
\label{eq10}
D=\{(u_1, \mu_1 ), (u_2, \mu_2 ),~\dots ~,(u_N, \mu_N)\}.
\end{equation}

using the training data, we can train a SVM using the below optimization problem:
\begin{subequations}
	\begin{flalign}
	&\min_{w, b, \epsilon} \dfrac{\textstyle 1}{\textstyle 2} w^T w + c \sum \epsilon_i, &\label{eq11a}\\
	subject~to:\nonumber \\
	&\mu_i (w^T \Phi (u_i)+ b ) \geq 1- \epsilon_i,  &\label{eq11b}\\
	&\epsilon_i \geq 1, &\label{eq11c}
	\end{flalign}
\end{subequations}

\noindent in which $2/w^Tw $ is the width of separating margin, $c$ is the regularization parameter, $\epsilon_i$ is the error in misclassifying user $i$, and $\Phi (u_i)$ is the Gaussian kernel function used to increase the precision of classification in problems with non-linearly separable data points, by capturing the correlations between different data points. It maps the features vector of each user $u_i$ into a point in higher dimensional transformed feature space. For any two m-dimensional feature vectors $u_i$ and $u_j$, the kernel function is defined as
\begin{equation}
\Phi _ {( u_{i}, u_{j} )}= e^ {-\gamma {|| u_{i}- u_{j} ||}^2}, \gamma= 1/2\sigma ^2 \geq 0.
\end{equation}

After deriving the classification vector $w$ and parameter $b$ from the optimization problem defined in (\ref{eq11a})-(\ref{eq11c}), we construct the classifier function
\begin{equation}
\label{eq13}
f(u_i)= (w^T u_i)+b = \begin{cases}
\geq 0,~i.e.~ \mu_i =1, \\
\le 0,~i.e.~\mu_i =0,
\end{cases}
\end{equation}

\noindent and use it to predict the association of each user $i$ with given feature vector $u_i$ as defined in (\ref{eq13}). Figure shows that for any potential user with the given six features, the SVM classifier function can determine if the MBS should offload it or serve it directly.
\begin{figure}
	\includegraphics[width=\linewidth]{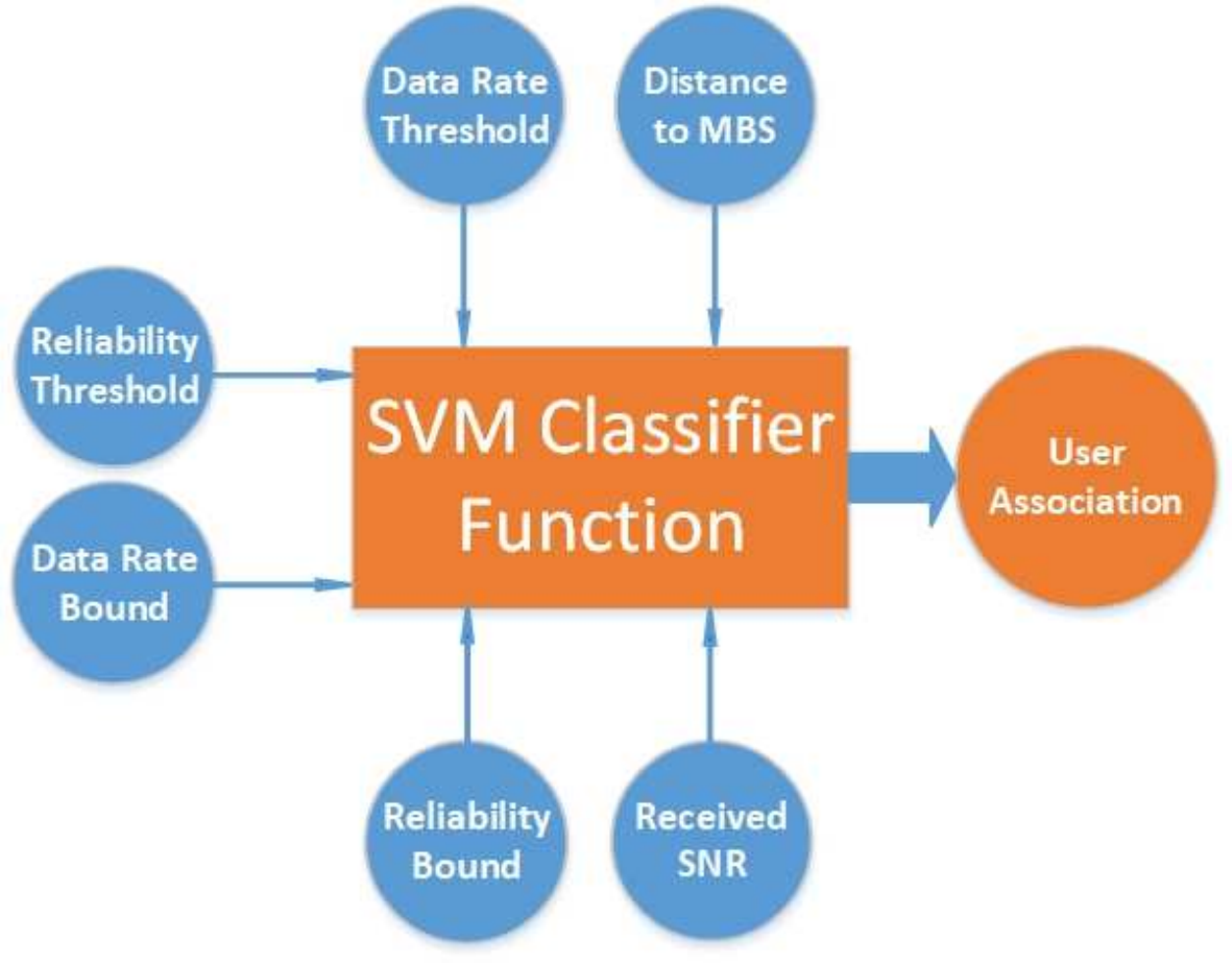}
	\centering
	\caption{SVM-based User Association in HetNets.}
	\label{fig:simulatedscenario}
\end{figure}

\subsection{Convex Resource Optimization (CRO)}
After determining the user associations using the SVM classifier by the MBS, the binary user association variables can be removed from the original JUR problem, and it can be reduced to a convex optimization problem. The non-convex objective function in JUR problem defined in (\ref{eq9a}) will be reduced to minimizing a linear cost function as defined in
\begin{equation}
\min_{p_i, w_i} \sum \limits_{i\in U,\mu_i=1} (c_p p_i + \gamma c_w w_i),
\label{eq14}
\end{equation}

which is a convex function in both power $p_i$, and bandwidth $w_i$ variables. Note that MBS is only optimizes its bandwidth and power allocations to those users that are not offloaded to SBSs, and have to be served by MBS. Also, the non-convex constraint defined in (\ref{eq9f}) can be removed since $\mu_i$ variables are no longer optimization variables, and are known to MBS using SVM classifier in previous phase. Since users who are associated to MBS experience the minimum delay which is the fixed computation delay of MBS, and none of them experience offloading delay, the delay constraint can also be removed in resource optimization problem for the MBS. Also, using the Markov's inequality bound for the reliability constraints, we have
\begin{equation}
Pr [r_{i} \geq r_{th,i}] \leq \frac{E[r_{i}]}{r_{th,i}},
\label{eq15}
\end{equation}

and, accordingly we can write  
\begin{equation}
Pr [r_{i} \leq r_{th,i}] = 1- Pr [r_{i} \geq r_{th,i}] \geq 1- \frac{E[r_{i}]}{r_{th,i}}.
\label{eq16}
\end{equation}

Hence, we can rewrite the reliability constraint defined in (\ref{eq9c}) as 
\begin{equation}
1- \frac{E[r_{i}]}{r_{th,i}} \leq \delta_{r,i}, 
\label{eq17}
\end{equation}

which can be simplified as
\begin{equation}
-E[r_{i}] + r_{th,i} (1-\delta_{r,i})\leq 0.
\label{eq18}
\end{equation}

It should be noted that according to (\ref{eq5}), knowing the transmission power, $p_i$ and bandwidth $w_i$, the expected service rate of user $i$, $E[r_i]$, is a function of expected channel gain and noise, and denoting expected channel gain and noise as $\bar{h}_i$ and $\bar{N}_0$, respectively, it can be calculated by
\begin{equation}
E[r_{i}]= w_i ~log (1+ p_i {\bar{h}_i}^2 /\bar{N}_0).
\label{eq19}
\end{equation}

Since $E[r_{i}]$ is a concave function in ($w_i$,$p_i$), hence $-E[r_{i}]$ and accordingly the reliability constraint defined in  (\ref{eq18}) are convex. The convex resource optimization (CRO) problem for MBS can be formulated as
\begin{subequations}
	\begin{flalign}
	&\min_{p_i, w_i} \sum \limits_{i\in U , \mu_i=1} (c_p p_i + \gamma c_w w_i), &\label{eq20a}\\\nonumber
	&~~~~~ subject~to: \nonumber &\\
	&-E[r_{i}] + r_{th,i} (1-\delta_{r,i})\leq 0, &\label{eq20b}\\
	& 0 \leq p_i \leq p_{max}, &\label{eq20c} \\
	& 0 \leq \sum \limits_{i\in U}w_i \leq W_{max}, &\label{eq20d} \\
	\nonumber
	\end{flalign}
\end{subequations}
\noindent which can be solved using CVX, in much less time than JUR problem. If some URLLC users have stricter delay constraints such that they cannot even wait for the computational delay,  $t_c$, of solving the CRO problem before receiving their service, then MBS has to find huristic methods to reduce the time complexity of solving this problem. One way to find the solution to the CRO optimization problem in less time is to use the method of Lagrange multipliers, since all its objective and constraints functions are differentiable and continious in both $p_i$ and $w_i$ optimization variables.

Defining the power vector $p=(p_1, p_2, \dots, p_N), ~0\leq p_i \leq P_{max},\forall i$, and the bandwidth vector $w=(w_1, w_2, \dots, w_N),~0 \leq \sum w_i \leq W_{max},\forall i$, and power and bandwidth cost function $f(p)=\sum_{i} c_p p_i $, and $f(w)= \sum_{i} \gamma c_w w_i $, the objective of the CRO problem is to find the optimal solution $(p^*, w^*)$ such that
\begin{equation}
(p^*, w^*)= min_{p,w} \{f(p) + f(w) |E[r_{i}] \geq r_{th,i} (1-\delta_{r,i}), \forall i \label{eq21} \}.
\end{equation}

The deviation of the offered reliability to each user $i$ and the minimum bound for the reliability of this user for any amount of allocated power, $p_i$, and bandwidth, $w_i$, is defined as
\begin{equation}
g(p_i,w_i)= E[r_{i}] - r_{th,i} (1-\delta_{r,i}), \forall i,
\label{eq22}
\end{equation} 

where, we must have $g(p_i,w_i) \geq 0$ to satisfy the reliability constraint of each user $i$ that is associated to the MBS. However, since $g(p_i,w_i)$ has a direct relation with both of the power and bandwidth cost functions, to minimize the cost we need to satisfy the reliability constraint of each user $i$ with minimum possible value for $g(p_i,w_i)$, which means that for the optimal solution, we want the this value to converge to zero. However, it is extremely important that the value of $g(p_i,w_i)$ stays positive while approaching to zero, since for negative values of it the reliability constraint of user $i$ will be violated. To guarantee this, we use a log barrier function for reliability in our penalized Lagrangian function for the method of multipliers to make sure that $g(p_i,w_i)$ will never turn into a negative value. By introducing the Lagrangian variable $\lambda_i$ for each user $i$ to model the cost of deviation from the required threshold for reliability, the penalized Lagrangian function for each user $i$, is given by 
\begin{equation}
L (p_i,w_i,\lambda_i )=f(p_i) +f(w_i) + \lambda_i ln~({g(p_i,w_i)}).
\label{eq23}
\end{equation}

Note that if $g(p_i,w_i) \leq 0$, then $ln~({g(p_i,w_i)})$ is undefined, hence $L (p_i,w_i,\lambda_i )$ can only be evaluated in the interior of the feasible region for reliability constraint. Denoting $\lambda = (\lambda_1, \lambda_2, \dots, \lambda_N)$ the Lagrangian function considering all the users will be given as
\begin{equation}
L (p,w,\lambda )= \sum_{\forall i \in U} L (p_i,w_i,\lambda_i ), \forall i \in U.
\label{eq24}
\end{equation} 

We denote the values of $p$, $w$, and $\lambda$ variables at each step $k$ of the Lagrangian method of multipliers as $p^k$, $w^k$, and $\lambda^k$, respectively. Starting from some initial values for these variables from their feasible regions, at each step $k$ we fix the values for two of these parameters in the Lagrangian function by using their current values, to find the optimal value for the third variable, by minimizing the Lagrangian function with respect to that variable. This update procedure is given in
\begin{equation}
\begin{cases}
\lambda^{k+1}= & Arg  \min_{\lambda}⁡{L (p^k,w^k,\lambda )}, \\
p^{k+1}= & Arg  \min_{p}⁡{L (p,w^k,\lambda^{k+1})}, \\
w^{k+1}= & Arg  \min_{w}⁡{L (p^{k+1},w,\lambda^{k+1} )} .\\
\end{cases}
\label{eq25}
\end{equation}

We continue this iterative updates until converging to a state in which the values of optimization variables do not change anymore. Note that the Lagrangian function is continuous and differentiable with respect to all three variables $p$, $w$, and $\lambda$, hence we can simply take a derivative from the Lagrangian function in each iteration to find its optimal value quickly w.r.t any variable when the values of other two variables are given. Using this method, in a few iterations we can find the optimal value for the Lagrange dual problem, and since the primal optimization problem is convex, the duality gap is zero, which means the optimal solution to the Lagrange dual problem is also the optimal solution to the primal optimization problem.

\section{Simulation Results}
\label{sec5}
To evaluate the efficiency of our proposed heuristic method, we consider a HetNet scenario in which there is one MBS located in the center of a cell with the radius of 2000 ft, and there are 8 overlaid SBSs with shorter coverage ranges of 600 ft within that cell who can serve the cellular users under their coverage. We also assume that there are 300 URLLC users who are randomly distributed within the cell, each with different data rate, reliability and delay constraints.
We solved the joint user association and resource allocation problem for the MBS using both JUR and LHM, by implementing these methods in Matlab. For better comparison, and to show the effects of user offloading on reducing the MBS's cost, we also implemented the Direct Serving Method (DSM) in which MBS serves all the users directly without offloading any of them to SBSs. In DSM, MBS optimizes its bandwidth and power allocations to minimize its serving cost using the cost function defined in (\ref{eq14}). We compare the performance parameters of these three methods in Table \ref{T1}.

\begin{center}
	\label{T1}
	\begin{tabular}{|c|c|c|c| }
		\hline
		Algorithm & DSM & JUR &  LHM	\\ [0.5 ex]
		\hline\hline
		Runing Time (sec)& 53.2131 & 54.0355 &	3.6809 \\
		\hline
		Avg Cost Per User & 102.6990 & 66.3524 &	68.6456\\
		\hline
		Serving Rate & 100\% & 100\% &	100\%\\
		\hline
		Total Offloaded Users & 0 & 226 &	222\\
		\hline
	\end{tabular}
\end{center}

As we can see from this table, in JUR method 226 users (75.33 \% of total 300 users), and in LHM 222 users (74 \% of total 300 users) have been offloaded to the overlaid SBSs, respectively which means that SVM classifier has successfully identified 98.23 \% of the users that must be offloaded in order to minimize the cost of MBS. Offloading of these users reduces the MBS's average energy and bandwidth cost per user from 102.6990 unit cost (UC) in DSM method to 66.35 UC and 68.64 UC in JUR and LHM methods, respectively which leads to the reduction of MBS's average energy and bandwidth consumption cost by nearly 36\%, 33\%, respectively as compared to DSM. As we can see in Table \ref{T1}, although the offloading rate and the average cost per user in both JUR and LHM are nearly the same, but the LHM's running time is much shorter than the time required by the JUR method. In fact, using the proposed learning based heuristic method to reduce the computational complexity of JUR, we reduced the resource allocation delay for the MBS from 54.03 sec in JUR (which is an NP-hard method) to only 3.68 sec in LHM, which leads to the 93\% reduction in spectrum access delay for URLLC users.

\begin{figure}
	\includegraphics[width=\linewidth]{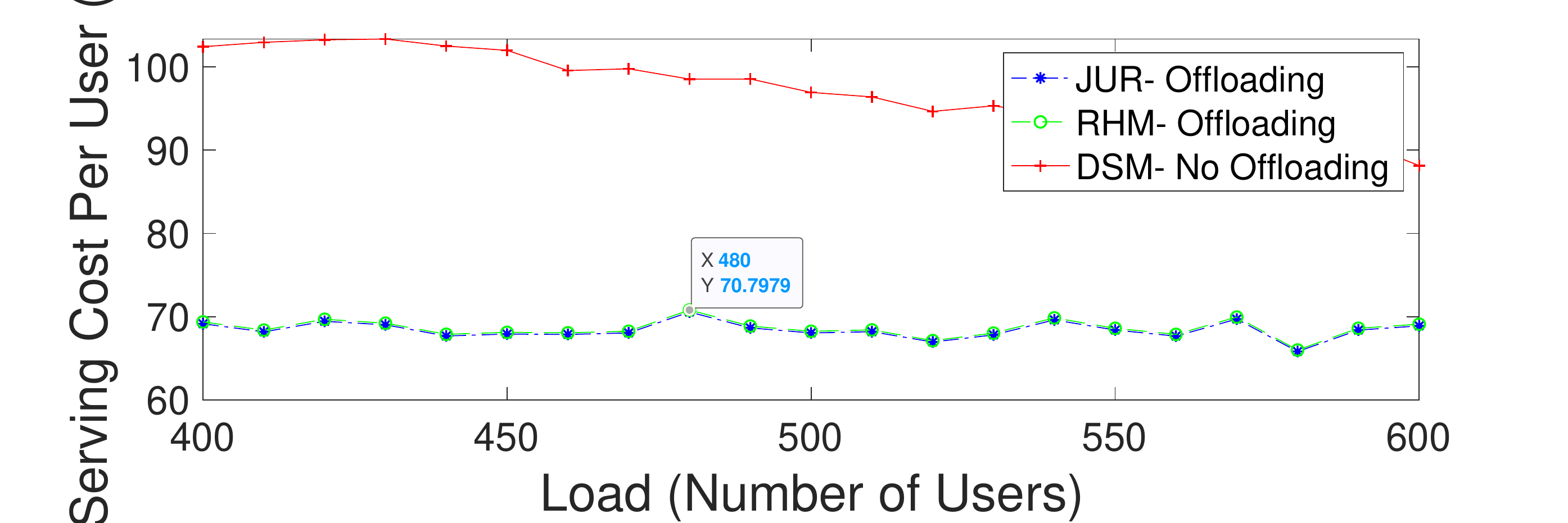}
	\centering
	\caption{Comparing serving cost per user in DSM, JUR and LHM.}
	\label{fig:cost comparison}
\end{figure}

Fig. \ref{fig:cost comparison} compares the MBS's average energy and bandwidth consumption cost to serve each user in JUR, LHM and DSM. As shown in this figure, due to the offloading of users from MBS to SBSs in LHM and JUR methods, the MBS's serving cost per user is much less in these methods as compared to DSM, and it is nearly the same in both LHM and JUR methods since the SVM classifier used in LHM heuristic has successfully identified and offloaded users from MBS to SBSs as in the JUR method with less than 2\% of users having different user-to-base station associations in LHM as compared to JUR. The reason for the gap between MBS's cost in serving users in LHM and JUR as compared to DSM is that SBSs usually have better channel conditions and hence consume less power and bandwidth to serve URLLC users under their coverage area, and offloading users located under the coverage area of SBSs has less cost for MBS as compared to serving them directly.

\begin{figure}
	\includegraphics[width=\linewidth]{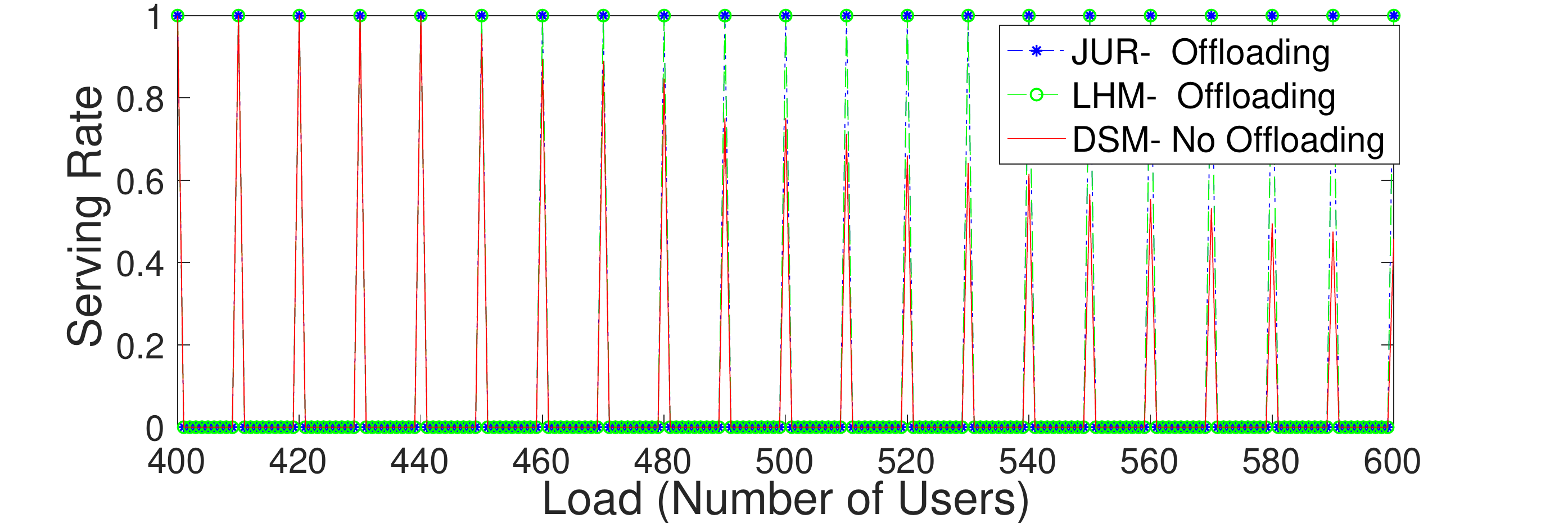}
	\centering
	\caption{Comparing service rates while increasing the load.}
	\label{fig:servicerates}
\end{figure}

To see the effects of load on the service rate of MBS using each of the JUR, LHM and DSM schemes, we change the number of users in our HetNet from 300 users to 500 users, by increasing the number of users with 20 new users in each step. We define the service rate as the percentage of URLLC users who are getting a service that satisfies their delay and reliability constraints. The Fig. \ref{fig:servicerates} compares the service rates of MBS using DSM, JUR, and LHM in different load situations. As we can see, by increasing the load or number of users in the HetNet, MBS is unable to serve all the users using DSM method, and the service rate goes below 50\% when the number of users exceeds 600 users, while in both JUR and LHM by exploiting the cooperation between MBS and SBSs, and offloading users to less congested SBSs, the full service rate can still be achieved as long as the load is less than the capacity of the HetNet.

\section{Conclusion}
\label{sec6}
In this paper, we have considered a HetNet model with one MBS, multiple SBSs and URLLC users with different latency and reliability constraints, to jointly optimize the user associations and resource allocation problems for the MBS. Modeling the latency and reliability constraints of users with probabilistic guarantees, we first formulated an optimization method for joint user association and resource allocation (JUR) in HetNets to minimize the cost for the MBS, and showed that it is NP-hard. In order to reduce the time complexity of the JUR method, we proposed a learning based heuristic method (LHM) to cast the initial optimization problem into a simple SVM-based user association model and a convex resource optimization (CRO) problem. To further reduce the delay for the users, we proposed an ADMM-based solution to the CRO problem. Simulation results validated the efficiency of the proposed method, and showed that it can reduce the MBS's energy and bandwidth consumption costs considerably by $\sim$ 33\%, while also reducing the spectrum access delay for cellular users by $\sim$ 93\% which is attractive to URLLC.




\begin{thebibliography}{1}
	
	\bibitem{Bennis:2018:URLLC}
	M. Bennis, M. Debbah and H. V. Poor, ``Ultrareliable and Low-Latency Wireless Communication: Tail, Risk, and Scale,'' in \emph{Proceedings of the IEEE,} vol. 106, no. 10, pp. 1834-1853, Oct. 2018.
	
%	\bibitem{Liu:2015:JUA}
%	D. Liu, Y. Chen, K. K. Chai, T. Zhang and K. Han, ``Joint user association and green energy allocation in HetNets with hybrid energy sources,'' \emph{ IEEE Wireless Communications and Networking Conference (WCNC)}, New Orleans, LA, pp. 1542-1547, 2015.
%	
%	\bibitem{Han:2017:BAU}
%	Q. Han, B. Yang, G. Miao, C. Chen, X. Wang and X. Guan, ``Backhaul-Aware User Association and Resource Allocation for Energy-Constrained HetNets,'' \emph{ IEEE Transactions on Vehicular Technology}, vol. 66, no. 1, pp. 580-593, Jan. 2017.
%	
%	
%	\bibitem{Borst:2013:ORA}
%	S. Borst, S. Hanly and P. Whiting, ``Optimal resource allocation in HetNets,'' \emph{ IEEE International Conference on Communications (ICC)}, Budapest, pp. 5437-5441, 2013.
%	
%	\bibitem{Hossain:2014:RAC}
%	Ekram Hossain; Long Bao Le; Dusit Niyato, ``Resource Allocation in CDMA-Based Multi-tier HetNets,'' \emph{ Radio Resource Management in Multi-Tier Cellular Wireless Networks}, Wiley Telecom, 2014, pp.352-, doi: 10.1002/9781118749821.ch8
%	
%	
%	
%	\bibitem{Lee:2018:MPS}
%	S. H. Lee and I. Sohn, ``Message-Passing Strategy for Joint User Association and Resource Blanking in HetNets,'' \emph{ IEEE Transactions on Wireless Communications}, vol. 17, no. 2, pp. 1026-1037, Feb. 2018.
	
	\bibitem{Chen:2017:DSA}
	L. Chen, B. Chang, G. Zhao and Z. Chen, ``Delay-sensitive area spectral efficiency optimization for uplink transmission in ultra-reliable and low-latency communications,'' \emph{23rd Asia-Pacific Conference on Communications (APCC)}, Perth, WA, pp. 1-6, 2017.
	
	
	\bibitem{Sun:2017:EER}
	C. Sun, C. She and C. Yang, ``Energy-Efficient Resource Allocation for Ultra-Reliable and Low-Latency Communications,'' \emph{IEEE Global Communications Conference (GLOBECOM)}, Singapore, pp. 1-6, 2017.
	
	
	\bibitem{Sutton:2018:EUR}
	G. J. Sutton et al., ``Enabling Ultra-Reliable and Low-Latency Communications through Unlicensed Spectrum,'' \emph{IEEE Network}, vol. 32, no. 2, pp. 70-77, March-April 2018.
	
	
	\bibitem{Kasgari: 2018: SOC}
	ATZ.  Kasgari  and  W.  Saad,  ``Stochastic  optimization  and control  framework  for  5G  network  slicing  with  effective  isolation,''  \emph{52nd Conference on Information Sciences and Systems (CISS)}, Princeton, USA, 2018.
	
	\bibitem{Munir:2017:ROM}
	H. Munir, S. A. Hassan, H. Pervaiz, Q. Ni and L. Musavian, ``Resource Optimization in Multi-Tier HetNets Exploiting Multi-Slope Path Loss Model,'' \emph{ IEEE Access}, vol. 5, pp. 8714-8726, 2017.
	
	\bibitem{Ji:2018:URL}
	H. Ji, S. Park, J. Yeo, Y. Kim, J. Lee and B. Shim, ``Ultra-Reliable and Low-Latency Communications in 5G Downlink: Physical Layer Aspects,'' \emph{ IEEE Wireless Communications}, vol. 25, no. 3, pp. 124-130, June 2018.
	
	\bibitem{Yousefvand:2017:DES}
	M. Yousefvand, T. Han, N. Ansari, A. Khreishah, ``Distributed Energy-Spectrum Trading in Green Cognitive Radio Cellular Networks,'' \emph{IEEE Trans. Green Commun. Netw.}, vol. 1, no. 3, pp. 253-263, Sept. 2017.
	
\end{thebibliography}
\end{document}